\documentstyle[11pt,epsf,epsfig]{article}
\newcommand{\NP}[1]{ Nucl.\ Phys.\ {#1}}

\newcommand{\PL}[1]{ Phys.\ Lett.\ { #1}}

\newcommand{\PR}[1]{Phys.\ Rev.\ { #1}}
\newcommand{\PRL}[1]{ Phys.\ Rev.\ Lett.\ { #1}}

\newcommand{\be}{\begin{equation}}
\newcommand{\bea}{\begin{eqnarray}}
\newcommand{\ee}{\end{equation}}
\newcommand{\eea}{\end{eqnarray}}

\begin{document}
%
{\large {\bf Chiral Perturbation Theory and Unitarization}\footnote{
Talk given by E.R.A. at the Miniworkshop on ``Few Quark Problems'', Bled,
Slovenia, July-2000.} }

{E. Ruiz Arriola$^{*}$, A. G\'omez Nicola$^{**}$, J. Nieves$^{*}$
and J. R. Pel\'aez$^{**}$}

{\footnotesize  $^{*}$ Departamento de F\'{\i}sica Moderna, Universidad
de Granada, E-18071 Granada, Spain.  \\ \phantom{----} $^{**}
$Departamento de F\'{\i}sica Te\'orica Universidad Complutense. 28040
Madrid, Spain.  }
\begin{abstract}
We review our recent work on unitarization and chiral perturbation
theory both in the $\pi\pi$ and the $\pi N$ sectors. We pay
particular attention to the Bethe-Salpeter and Inverse Amplitude
unitarization methods and their recent applications to  $\pi\pi$
 and $\pi N$ scattering.
\end{abstract}

\section{Introduction}

 Chiral
Perturbation Theory (ChPT) is
 a practical and widely accepted effective field theory to deal with low
energy processes in hadronic physics.
 \cite{GL84,Le94,Reviews}.
The essential point stressed in this approach is that the low
energy physics does not depend on the details of the short
distance dynamics, but rather on some bulk properties effectively
encoded in the low energy parameters. This point of view has been
implicitly adopted in practice in everyday quantum physics; well
separated energy and distance scales can be studied independently
of each other. The effective field theory approach also makes such
a natural idea into a workable and systematic computational
scheme.

The great advantage of ChPT is that the  expansion parameter can
be clearly identified {\it a priori}
 (see e.g. Ref.~\cite{Reviews}) when carrying out
systematic calculations of mass splittings, form factors and
scattering amplitudes. However, the connection to the underlying
QCD dynamics becomes obscure since the problem is naturally
formulated in terms of the relevant hadronic low energy degrees of
freedom with no explicit reference to the fundamental quarks and
gluons.
In addition, in some cases (see below) a possible drawback is the
lack of numerical convergence of such an expansion when confronted
to experimental data, a problem that gets worse as the energy of
the process increases. Recent analysis  provide good examples of
both rapid convergence and slow convergence in ChPT. In the
$\pi\pi$ sector the situation to two loops~\cite{pipi2} seems to
be very good for the scattering lengths. Here, the expansion
parameter is $m_\pi^2 /(4\pi f_\pi)^2 = 0.01 $ ($m_\pi=139.6 {\rm
MeV}$ the physical mass of the charged pion) and the coefficients
of the expansion
 are of order unity. For instance, to two loops (third
order in the expansion parameter) the expansion of the s-wave of
the isospin $I=0$ channel reads~\cite{ej.epj}
\begin{eqnarray}
a_{00} m_\pi &=& \underbrace{0.156}_{\rm tree} + \underbrace{
0.043 \pm 0.003}_{\rm 1 \, loop} + \underbrace{0.015 \pm
0.003}_{\rm 2 \, loops} + \cdots
\end{eqnarray}
where the theoretical errors are described in ~\cite{ej.epj}.
 Thus the expansion up to two loops is both convergent $ a^{(n)}
>> a^{(n+1)} $ and predictive $ \Delta a^{(n)} << a^{(n+1)} $ , $
\Delta a^{(n)} << a^{(n)} $. The {\it prediction} of ChPT of
s-wave scattering lengths $a_{IJ} $ in the isospin $I=0$ and $I=2$
channels
 yields ~\cite{ej.epj}
\begin{eqnarray}
a_{0 \, 0} m_\pi &=& +0.214 \pm 0.005 \quad ({\rm exp.} \, +0.26
\pm 0.05 ) \nonumber \\ a_{2 \, 0} m_\pi &=& -0.420 \pm 0.010
\quad ({\rm exp.} \, -0.28 \pm 0.12 )
\end{eqnarray}
The theoretical predictions for these observables are an order of
magnitude more accurate than the corresponding experimental
numbers. Note that in hadronic physics we are usually dealing with
the opposite situation, confronting accurate measurements to
inaccurate theoretical model calculations. On the contrary, for
$\pi N$ scattering, for ChPT in the heavy baryon formulation
 the expansion is less rapidly converging than in the
$\pi\pi$ case, since NLO corrections become comparable to the LO
ones. For instance in the $P_{33}$-channel the expansion up to
third order, after a fit to the threshold properties,
reads~\cite{Mo98}
\begin{eqnarray}
a_{33}^1 m_\pi^3 = \underbrace{35.3}_{\rm 1st \,  order} +
\underbrace{47.95}_{\rm 2nd \, order} - \underbrace{1.49}_{\rm 3rd
\, order} + \cdots = 81.8 \pm 0.9 \qquad ( {\rm exp.} 80.3 \pm 0.6
)
\end{eqnarray}
Here first order means $1/f_\pi^2$, second order $1/f_\pi^2 M_N $
and third order $1/f_\pi^4$ and $1/f_\pi^2 M_N^2 $. Despite these
caveats, there is no doubt that the effective field theory
approach provides a general framework where one can either verify
or falsify, not only bulk properties of the underlying dynamics,
but also the dynamics of all models sharing the same general
symmetries of QCD.

Finally, let us remark that  the perturbative nature of the chiral
expansion makes the generation of pole singularities, either bound
states or resonances, impossible from the very begining unless
they are already present at lowest order in the expansion. There
are no bound states in the $\pi\pi$ and $\pi N$ systems, but the
$\rho$ and the $\Delta$ resonances are outstanding features of
these reactions dominating the corresponding cross sections at the
C.M. energies $\sqrt{s} = m_\rho=770 {\rm MeV} $ and $\sqrt{s} =
m_\Delta = 1232 {\rm MeV}$ respectively.

\section{The role played by unitarity}

Exact Unitarity plays a crucial role in the description of
resonances. However, ChPT only satisfies unitarity perturbatively.
Nevertheless, there are many ways to restore exact unitarity out
of perturbative information, i.e.: the $K$-matrix method
Ref.~\cite{K-matrix}, the Inverse Amplitude Method (IAM)
Ref.~\cite{IAM,IAMcoupled}, the Bethe-Salpeter Equation (BSE)
Ref.~\cite{ej.plb,ej.npa}, the N/D method Ref.~\cite{N/D}, etc. (
See e.g. Ref.~\cite{OOR2000} for a recent review), which are
closely related to one another.

Some of these Unitarization methods have been very successful
describing experimental  data in the intermediate energy region
including the resonant behavior.  Despite this success the main
drawback is that this approach is not as systematic as standard
ChPT, for instance in the estimation of the order of the neglected
corrections. In this work we report on our most recent works
related to unitarization, both in the $\pi\pi$ and $\pi N$
sectors, where we have obtained intermediate energy {\it
predictions} using the chiral parameters and their error bars
obtained from standard ChPT applied at low energies. This means in
practice transporting the possible correlations among the fitting
parameters obtained from a fit where the phase shifts are assumed
to be gauss-distributed in an uncorrelated way.

\section{Results in the $\pi\pi$ sector}

The ChPT expansion displays a very good convergence in the
meson-meson sector. As a consequence the unitarization methods
have been very successful in extending the ChPT applicability to
higher energies. In particular, within a coupled channel IAM
formalism, it has been possible to describe all the meson-meson
scattering data, even the resonant behavior, below 1.2 GeV
\cite{IAMcoupled}, but without including explicitly any resonance
field. These works have already been extensively described in the
literature  and here we will concentrate in the most recent works
of two of the authors on the $\pi\pi$ sector, dealing with the
Bethe-Salpeter equation, which has the nice advantage of allowing
us to identify the diagrams which are resumed.

\begin{figure}[t]
\vspace{-10.5cm}\hspace*{-1cm} \hbox
to\hsize{\hfill\epsfxsize=0.75\hsize \epsffile[52 35 513
507]{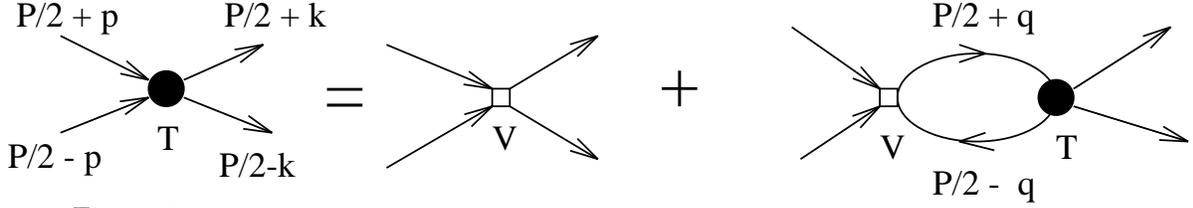}\hfill} \vspace{0.5cm}
\caption[pepe]{\footnotesize Diagrammatic representation of the
BSE equation. It is also sketched the used kinematics.}
\vspace{0.5cm} \label{fig:kin}
\end{figure}

Indeed, any unitarization method performs, in some way or another,
an infinite sum of perturbative contributions. At first sight,
this may seem arbitrary, but some constraints have to be imposed
on the unitarization method to comply with the spirit of the
perturbative expansion we want to enforce. In the BSE approach the
natural objects to be expanded are the {\it potential} and the
{\it propagators}. The BSE as it has been used in
Refs.~\cite{ej.plb,ej.npa} reads (See Fig.~\ref{fig:kin})
\begin{eqnarray}
T_P^I (p,k) &=& V_P^I (p,k) + {\rm i}\int\frac{d^4 q}{(2\pi)^4}T_P^I
(q,k)\Delta(q_+) \Delta(q_-) V_P^I (p,q)\label{eq:bs}
\end{eqnarray}
where $q_{\pm} = (P/2\pm q)$ and $T_P^I (p,k)$ and $V_P^I (p,k)$
are the total scattering amplitude\footnote{The normalization of
the amplitude $T$ is determined by its relation with the
differential cross section in the CM system of the two identical
mesons and it is given by $d\sigma /d\Omega = |T_P(p,k)|^2 /
64\pi^2 s$, where $s=P^2$. The phase of the amplitude $T$ is such
that the optical theorem reads ${\rm Im} T_P(p,p) = - \sigma_{\rm
tot} (s^2-4s\,m^2)^{1/2}$, with $\sigma_{\rm tot} $ the total
cross section. The contribution to the amputated Feynman diagram
is $(-{\rm i} T_P(p,k) )$ in Fig.~\protect\ref{fig:kin}.} and
potential for the channel with total isospin $I=0,1,2$.
and then the projection over each partial wave $J$ in the CM frame,
$T_{IJ}(s)$, is given by
\begin{eqnarray}
T_{IJ}(s) = \frac12 \int_{-1}^{+1}
P_J\left(\cos\theta\right) T_P^I(p,k)~
d(\cos\theta) &=& \frac{{\rm i} 8\pi
s}{\lambda^{\frac12}(s,m^2,m^2)}
\left [ e^{2{\rm i}\delta_{IJ}(s)} -1 \right ]
\end{eqnarray}
where $\theta$ is the angle between $\vec{p}$ and $\vec{k}$ in the CM
frame, $P_J$ the Legendre polynomials and $\lambda(x,y,z) =
x^2+y^2+z^2 -2xy-2xz-2yz$. Notice that in our normalization the
unitarity limit implies $ |T_{IJ}(s)| < 16 \pi s / \lambda^{1/2}
(s,m^2,m^2) $.

\begin{figure}
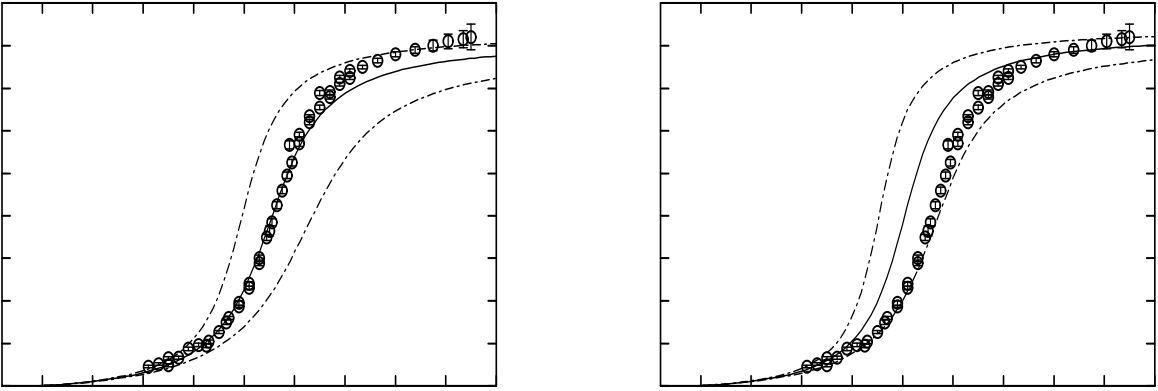

\hspace*{-.8cm} \centerline{
\epsfig{figure=pha_shifts.ps3,height=6cm,width=7.5cm} \hspace{1cm}
\epsfig{figure=pha_shifts.ps4,height=6cm,width=7.5cm}}
\caption[pepe]{\footnotesize $ I=J=1$ $\pi\pi$ phase shifts as a
function of the total CM energy $\protect\sqrt s$ for both sets of
${\bar l}$'s given in Eq.~(\protect\ref{eq:elesab}). Left (right)
 figures have been obtained with the set {\bf A} ({\bf B}) of
parameters. Solid lines are the predictions of the off-shell BSE
approach, at lowest order, for the different $IJ-$channels. Dashed
lines are the 68\% confidence limits. Circles stand for the
experimental analysis of Refs.~\protect\cite{pa73} and
~\protect\cite{em74}.}
\label{fig:off-res}
\end{figure}

The solution of the BSE at lowest order, i.e., taking the free
propagators for the mesons and the potential as the tree level
amplitude can be obtained after algebraic manipulations described in
detail in Refs.~\cite{ej.plb,ej.npa}, renormalization and matching to
the Taylor expansion up to second order in $s-4m^2$ of the one loop
chiral perturbation theory result. We only quote here the result for
the $\rho$ channel:
\begin{eqnarray}
T^{-1}_{11}(s) &=& -\bar I_0(s) + {1\over 16\pi^2} \left[ 2 ( \bar l_2
- \bar l_1 ) + {97\over 60} \right] \\ &+& {1\over s-4m^2} \left\{
{m^2 \over 4 \pi^2 } \left[ 2 ( \bar l_2 - \bar l_1 ) + 3 \bar l_4 -
{65 \over 24} \right] - 6 f^2 \right\}
\end{eqnarray}
where the unitarity integral $\bar I_0 (s) $ reads
\begin{eqnarray}
{\bar I}_0 (s) \equiv I_0(s)-I_0(4 m^2 ) &=& \frac{1}{(4\pi)^2}
\sqrt{1-\frac{4m^2}{s}} \log \frac{\sqrt{1-\frac{4m^2}{s}}+1
}{\sqrt{1-\frac{4m^2}{s}}-1}
\end{eqnarray}
Here the complex phase of the argument of the $\log$ is taken in the
interval $[-\pi,\pi[$. Similar expressions hold for the
scalar-isoscalar ($\sigma$) and the scalar-isotensor channels. Notice
that in this, so-called {\it off-shell} scheme, the left hand cut is
replaced at lowest order by a pole in the region $ s << 0 $.  For the
low energy coefficients $\bar l_{1,2,3,4} $ we take the values
\begin{eqnarray}
{\rm set\,\, {\bf A}:\,\,\,} && {\bar l}_1 = -0.62 \pm
0.94,\,\,{\bar l}_2 = 6.28 \pm 0.48,\,\, {\bar l}_3 = 2.9 \pm 2.4,\,\,
{\bar l}_4 = 4.4 \pm 0.3 \nonumber\\
{\rm set\,\, {\bf B}:\,\,\,} && {\bar l}_1 = -1.7\phantom{0} \pm
1.0\phantom{0},\,\,{\bar l}_2 = 6.1\phantom{0} \pm 0.5\phantom{0},\,\,
{\bar l}_3 = 2.9 \pm 2.4,\,\, {\bar l}_4 = 4.4 \pm 0.3
\label{eq:elesab}
\end{eqnarray}
In both sets ${\bar l}_3$ and ${\bar l}_4$ have been determined from
the $SU(3)$ mass formulae and the scalar radius as suggested
in~\cite{GL84} and in~\cite{bct98}, respectively. On the other hand
the values of ${\bar l}_{1,2}$ come from the analysis of
Ref.~\cite{rdgh91} of the data on $K_{l4}-$decays (set {\bf A}) and
from the combined study of $K_{l4}-$decays and $\pi\pi$ with some
unitarization procedure (set {\bf B}) performed in Ref.~\cite{bcg94}.
The results for the $\rho$ channel are shown  in
Fig.~\ref{fig:off-res}.

As discussed in Ref.~\cite{ej.npa} it is also possible to take into
account the left hand cut in the so-called {\it on-shell} scheme where
it can be shown that after renormalization the on-shell unitarized
amplitude acquires the following form
\begin{eqnarray}
T_{IJ} (s)^{-1} +\bar{I}_0 (s) - V_{IJ} (s)^{-1}  &=&
T_{IJ} (s_0)^{-1} +\bar{I}_0 (s_0) - V_{IJ} (s_0)^{-1}  = -C_{IJ}
\end{eqnarray}
where $C_{IJ}$ should be a constant, independent of $s$ and the
subtraction point $s_0$, and chosen to have a well defined limit when
$m\to 0$ and $1/f \to 0$. $V_{IJ} (s) $ is the {\it
on-shell}-potential and has the important property of being real for
$0 < s <16m^2 $, and presenting cuts in the four pion threshold and the
left hand cut caused by the unitarity cuts in the $t$ and $u$
channels. The potential can be determined by matching the amplitude to
the ChPT amplitude in a perturbative expansion. This method provides a
way of generating a unitarized amplitude directly in terms of the low
energy coefficients $\bar l_{1,2,3,4} $ and their errors. In this {\it
on-shell} scheme a successful description of both $\pi\pi$ scattering
data as well as the electromagnetic pion form factor, in agreement with
Watson's theorem, becomes possible yielding a very accurate
determination of some low energy parameters. The procedure to do this
becomes a bit involved and we refer to Ref.~\cite{ej.npa} for further
details. We should also say that the one-loop unitarized amplitudes
generate the complete ChPT result and {\it some} of the two and higher
loop results. The comparison of the generated two-loop contribution of
threshold parameters with those obtained from the full two loop
calculation is quantitatively satisfactory within uncertainties.

\section{Results in the $\pi N$ sector}

The methods and results found in the $\pi\pi$ system are very
encouraging, suggesting the extension to the $\pi N$ system.
However, ChPT does not work in the $\pi N $ sector as nicely as it
does in the $\pi\pi$ sector. As we will see, the low convergence
rate of the chiral expansion makes it difficult to match standard
amplitudes to unitarized ones in a numerically sensible manner.
After an initial attempt within the relativistic
formulation~\cite{gss88}, it was proposed to treat the baryon as a
heavy particle well below the nucleon production
threshold~\cite{jm91}. The resulting Heavy Baryon Chiral
Perturbation Theory (HBChPT)  provides a consistent framework for
the one nucleon sector,
 particularly in $\pi N $ scattering ~\cite{Mo98}. The proposal of
Ref.~\cite{bl99} adopting the original relativistic formalism but
with a clever renormalization scheme seems rather promising but
unfortunately the phenomenological applications to $\pi N $ scattering
 have not been worked out yet.

\subsection{The IAM method in $\pi$-N scattering}

The inverse amplitude method (IAM) is a unitarization method where
the inverse amplitude, and not the amplitude, is expanded, i.e.,
if we have the perturbative expansion for the partial wave
amplitude $f(\omega)$,
\begin{eqnarray}
f(\omega) = f_1 (\omega)  + f_2 (\omega)  + f_3 (\omega) + \dots
\end{eqnarray}
with $\omega=\sqrt{q^2+m_\pi^2}$ and $q$ the
C.M. momentum, then one considers the expansion
\begin{eqnarray}
{1\over f(\omega)} = {1\over f_1 (\omega) } - {f_2(\omega) +
f_3(\omega) \over [ f_1 (\omega)]^2} + { f_2 (\omega)^2 \over
[ f_1 (\omega)]^3 } + \dots
\end{eqnarray}
The IAM fulfills exact unitarity, ${\rm Im} f(\omega)^{-1} = -q$ and
reproduces the perturbative expansion to the desired order.

\begin{figure}[htbp]
\begin{center}
\leavevmode \epsfysize = 215pt
\makebox[0cm]{\epsfbox{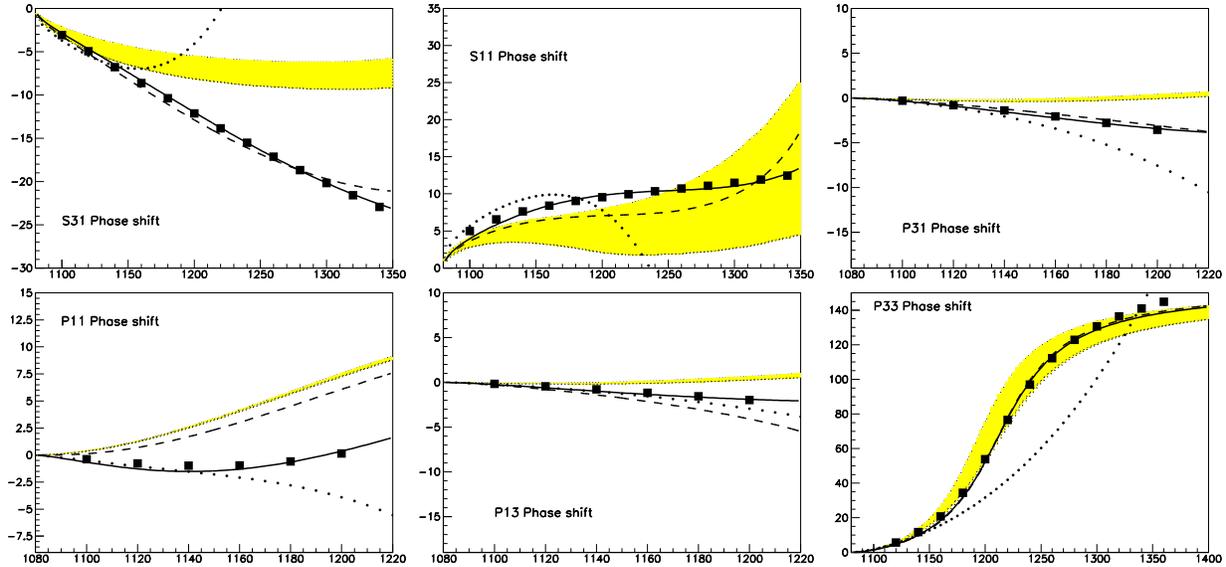}}
\end{center}
\caption[pepe] {\footnotesize Phase shifts in the IAM method as a
function of the C.M. energy $\sqrt{s}$. The shaded area
corresponds to the result of propagating the errors of the chiral
parameters obtained from low-energy data (see
Ref.~\cite{gnpr2000}). This illustrates the uncertainties due to
the choice of different parameter sets from the literature. The
dotted line is the extrapolated HBChPT result. The continuous line
is an unconstrained IAM {\it fis} to the data, whereas for the
dashed line the fit has been constrained to the resonance
saturation hypothesis.  } \label{fig:iam}
\end{figure}

This
method has been applied for $\pi N$ scattering~\cite{gp99} to
unitarize the HBChPT results of Ref.~\cite{Mo98} to third order.
In this context, it is worth pointing out that the use of a similar
 unitarization method, together with very simple phenomenological
 models, was already  successfully undertaken in the 70's
 (see ~\cite{70s} and references therein). Nonetheless, a
systematic application within an effective Lagrangian approach
 was not carried out.
In  ~\cite{gp99} the  phase shifts for the partial waves up to the
inelastic thresholds have been fitted, obtaining the right pole
for the $\Delta (1232)$ in the $P_{33}$ channel.
 In that work, it has been pointed out that   to get the best accuracy
 with data, one needs chiral parameters  of unnatural size,
very different from those of perturbative HBChPT. This is most likely
related  to the slow convergence rate of the expansion. However,
 it must be stressed that one can still reproduce the $\Delta (1232)$
 with second order parameters compatible with
the hypothesis of resonance saturation ~\cite{BeKaMe}. In a
subsequent work~\cite{ejweb} we have proposed an improved IAM
method based on a reordering of the HBChPT series. The encouraging
results for the $\Delta$-channel have also been extended  to the
remaining low partial waves~\cite{gnpr2000}, as it can be seen in
Fig.~\ref{fig:iam}. In this case, the size of the
 chiral parameters is natural and the
$\chi^2$ per d.o.f is considerably better than
 the IAM applied to plain HBChPT.

\subsection{BSE method and the $\Delta$-resonance}

\begin{figure}
\vspace*{-1.cm}
\begin{center}
\leavevmode \epsfysize = 600pt
\makebox[0cm]{\epsfbox{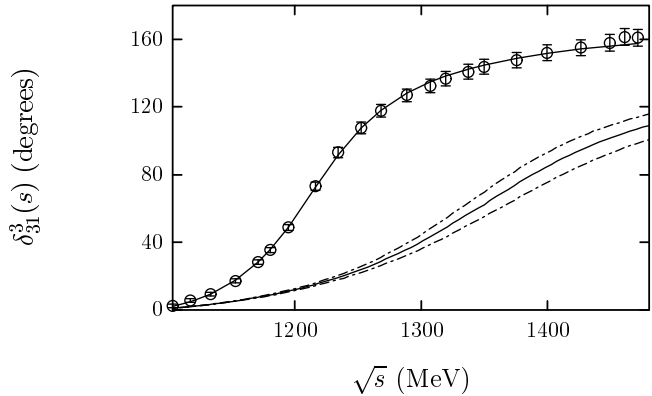}}
\end{center}
\vspace{-15.cm} \caption[pepe]{\footnotesize $P_{33}$ phase shifts
as a function of the total CM energy $\protect\sqrt s$. The upper
solid line represents a $\chi^2$-fit of the parameters,
$c_{0,1,2,3}$ to the data of Ref.~\protect\cite{AS95} (circles).
Best fit parameters are denoted $c_{0,1,2,3}^{\rm fit}$ in the
main text. The lower lines stand for the results obtained with the
parameters deduced from HBChPT and denoted $c_{0,1,2,3}^{\rm th}$.
Central values lead to the solid line, whereas the errors on
$c_{0,1,2,3}^{\rm th}$ lead to the dash-dotted lines.}
\label{fig:delta}
\end{figure}

Recently~\cite{ej.prd}, we have used the BSE to HBChPT at lowest
order in the chiral expansion and have looked at the $P_{33}$
channel. We have found a dispersive solution which needs four
subtraction constants,
\begin{eqnarray}
f^{3/2}_{3/2\,,1} (\omega)^{-1} &=& -\frac{ 24 \pi}{ (\omega^2 - m^2)}
\left\{ \frac{-f^2 \omega}{2 g_A^2} + P(\omega) +
(\omega^2-m^2)\bar{J}_0(\omega)/6 \right\} \nonumber\\ &&\nonumber\\
P(\omega) &=& m^3\left ( c_0 + c_1 (\frac{\omega}{m}-1) + c_2
(\frac{\omega}{m}-1)^2 + c_3 (\frac{\omega}{m}-1)^3 \right )
\end{eqnarray}
where the unitarity integral is given by
\begin{eqnarray}
\bar{J}_0 (\omega) \equiv J_0 (\omega) - J_0 (m) &=&
-\frac{\sqrt{\omega^2-m^2}}{4\pi^2} \{ {\rm arcosh} \frac{\omega}{m}
-{\rm i} \pi\}; \qquad \omega > m
\end{eqnarray}
The $\chi^2 $ fit yields the following numerical values for the parameters:
\begin{eqnarray}
c_0^{\rm fit}  = 0.045 \pm 0.021 \, , \quad  c_1^{\rm fit}  = 0.29 \pm 0.08
\, , \quad  c_2^{\rm fit}  = -0.17 \pm 0.09 \, , \quad  c_3^{\rm fit} = 0.16 \pm 0.03
\nonumber
\end{eqnarray}
with $ \chi^2/{\rm d.o.f.} = 0.2 $. However, if we match the
coefficients with those stemming from HBChPT we would get instead
the following numerical values:
\begin{eqnarray}
c_0^{\rm th}   = 0.001 \pm 0.003 \, , \quad
c_1^{\rm th}  = 0.038 \pm 0.006  \, , \quad
c_2^{\rm th}  = 0.064 \pm 0.005  \, , \quad
c_3^{\rm th} = 0.036 \pm 0.002 \nonumber
\end{eqnarray}
The discrepancy is, again, attributed to the low convergence rate of
the expansion. The results for the $P_{33}$ phase shift both for the
fit and the MonteCarlo propagated errors of the HBChPT matched
amplitudes have been depicted in Fig.~\ref{fig:delta}

\subsection{$\pi N$ scattering and the $N^* (1535) $ resonance}

\begin{figure}[]
\vspace*{-1cm}
\begin{center}
\epsfysize = 500pt
\makebox[0cm]{\epsfbox{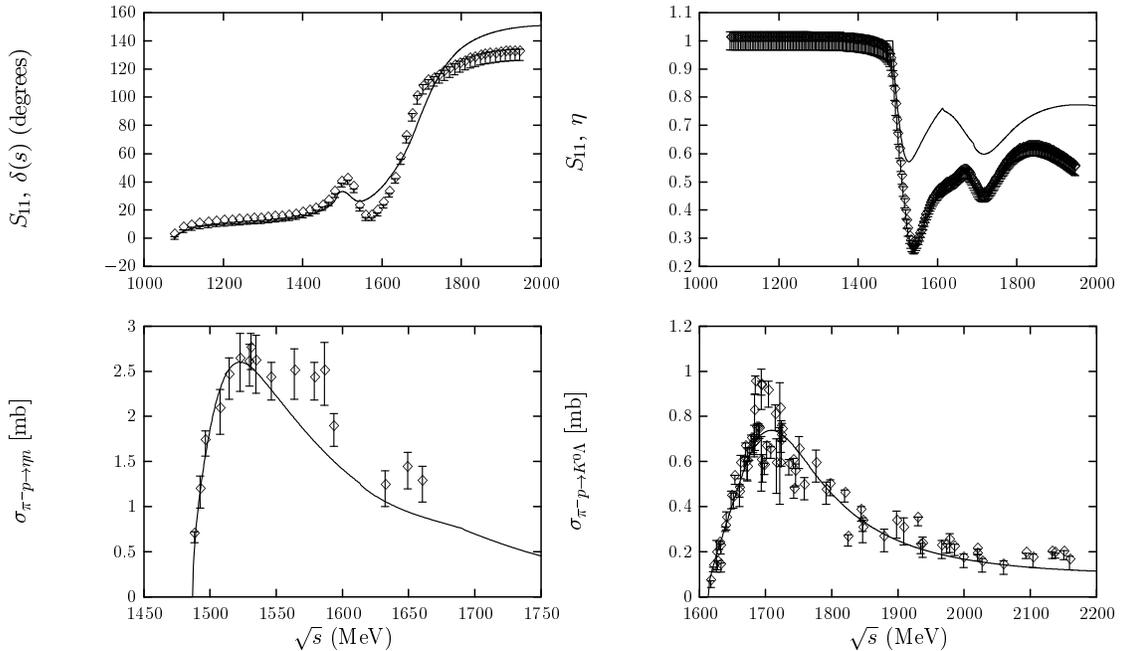}}
\end{center}
\vspace{-8.3cm} \caption[pepe] {\footnotesize $\pi N $ scattering
BSE results as a function of C.M. energy $\sqrt{s}$. Upper left
figure: $S_{11}$ phase shifts. Upper right figure: inelasticity in
the $\pi N$ channel. Lower left figure: $\pi N \to \eta N $ cross
section. Lower right figure: $\pi N \to K \Lambda $ cross section.
Data from Ref.~\cite{AS95}.  (See Ref.~\cite{ej2001} for further
details.)} \label{fig:acoplado}
\end{figure}

One of the  greatest advantages of both the IAM and the BSE methods
 is that the generalization to
include coupled channels is rather straightforward. For the IAM
case we refer to \cite{IAMcoupled} for more details. As for the
BSE, two of the authors~\cite{ej2001} have dealt with the problem
in the $S_{11}$ channel, at $\sqrt{s}$ up to $1800 {\rm GeV}$
using the full relativistic, rather than the heavy baryon
formulation used in Ref.~\cite{ksw95} for the s-wave and extended
in Ref.~\cite{Ca2000} to account for p-wave effects. For these
energies there are four open channels, namely $\pi N $,$\eta N
$,$K \Sigma $ and $K \Lambda $, so that the BSE becomes a $4
\times 4 $ matrix equation for the $I=1/2$, $J=1/2$ and $L=0$
partial wave. Additional complications arise due to the Dirac
spinor structure of the nucleon, but the BSE can be analytically
solved after using the above mentioned off-shell renormalization
scheme. It turns out~\cite{ej2001} that to lowest order in the
potential and the propagators, one needs 12 unknown parameters,
which should be used to fit experimental data. Several features
make the fitting procedure a bit cumbersome. In the first place,
there is no conventional analysis in the relativistic version of
ChPT for this process and thus no clear constraints can be imposed
on the unknown parameters. Secondly, the channel $\pi N\to \pi \pi
N $ is not included in our calculation. Therefore one should not
expect perfect agreement with experiment, particularly in the
elastic channel since it is known that $10-20 \% $ of the $N^* $
resonance decay width goes into $\pi\pi N$. On the other hand, one
cannot deduce from here how important is the $\pi\pi N $ channel
in the $\eta$ production channel, $\pi N \to \eta N $. Actually,
in Ref.~\cite{Na2000} it has been suggested that the bulk of the
process may be explained without appealing to the three body
intermediate state $\pi \pi N$. The work of Ref.~\cite{Na2000}
would correspond in our nomenclature to the on-shell lowest order
BSE approximation, which by our own experience describes well the
bulk of the data. With the BSE we have provided a further
improvement at low energies by including higher order corrections.
In the absence of a canonical low energy analysis it seems wiser
to proceed using the off-shell renormalization scheme. In
Fig.~\ref{fig:acoplado} we present a possible 12 parameter fit
which accounts both for the elastic low and intermediate energy
region and the lowest production channels. The failure to describe
data around the $N^*$ resonance is expected, since as we have
already mentioned the $\pi\pi N$ channel must be included. Our
results seem to confirm the assumption made in Ref.~\cite{Na2000}
regarding the unimportance of the three body channel in describing
the coupling of the $N^*$ resonance to $\eta N$.

\section{Conclusions and Outlook}

The results presented here show the success and provide further
support for unitarization methods complemented with standard
chiral perturbation theory, particularly in the case when
resonances are present. But unitarization by itself is not a
guarantee of success; the unitarization method has to be carefully
chosen so that it provides a systematic convergent and predictive
expansion, as we have discussed above. In order to describe the
data in the intermediate energy region the chiral parameters can
then be obtained from
\begin{itemize}
\item Either a direct $\chi^2$ fit of the order by order unitarized amplitude
and the corresponding low energy parameters. The upper energy limit is
determined by imposing an acceptable description $\chi^2 /{\rm DOF}
\sim 1 $.
\item Or from a low energy determination of the low energy parameters with errors
by performing a $\chi^2$-fit of the standard ChPT amplitude until $\chi^2
/{\rm DOF}\sim 1 $, and subsequent MonteCarlo error propagation of the
unitarized amplitude.
\end{itemize}
Differences in the low energy parameters within several methods should
 be compatible within errors, as long as  the Chiral expansion  has a
 good convergence. But,
unfortunately this is not always the case. Clearly, the $\pi N$
sector is not only more cumbersome theoretically than the $\pi\pi$
sector but also more troublesome from a numerical point of view.
Standard ChPT to a given order can be seen as a particular choice
which sets higher order terms to zero in order to comply with
exact crossing but breaking exact unitarity. The unitarization of
a the ChPT amplitude is also another choice of higher order terms
designed to reproduce exact unitarity but breaking exact crossing
symmetry. Given our inability to write a closed analytic expresion
for an amplitude in a chiral expansion which simultaneously
fulfills both exact crossing and unitarity we have preferred exact
unitarity. This is justified a posteriori by the
 successful description of data in the
intermediate energy region, which indeed
 suggests a larger convergence radius of the
chiral expansion.

\section*{Acknowledgments}
This research was partially supported by DGES under contracts
PB98--1367 and by the Junta de Andaluc\'{\i}a FQM0225 as well as by
DGICYT under contrascts AEN97-1693 and PB98-0782.


\end{document}